\documentclass[a4paper, 11pt]{article}
\usepackage[english]{babel}
\usepackage{units}
\usepackage{setspace}
\usepackage{graphicx}
\usepackage{float}
\usepackage{flafter}
\usepackage{times}
\usepackage{enumerate}
\usepackage{url}
\usepackage{amsmath}
\usepackage{textcomp}
\usepackage[square,numbers,sort&compress]{natbib}
\usepackage{longtable}
\usepackage{comment}
\usepackage{placeins}
\usepackage[version=3]{mhchem} 
\usepackage{braket}
\usepackage{amsfonts}
\usepackage[superscript]{cite} 

\usepackage[T1]{fontenc}
\usepackage{blindtext}
\usepackage[labelfont=bf]{caption}

\title{Graph Convolutional Neural Networks for (QM)ML/MM Molecular Dynamics Simulations}

\author{Albert Hofstetter,
        Lennard B\"oselt,
        Sereina Riniker$^\text{*}$}

\date{\small [*] \textit{Laboratory of Physical Chemistry, ETH Zurich, Vladimir-Prelog-Weg 2, 8093 Zurich, Switzerland \\Email: sriniker@ethz.ch}}

\begin{document}

\maketitle

\section*{Abstract}
To accurately study chemical reactions in the condensed phase or within enzymes, both a quantum-mechanical description and sufficient configurational sampling is required to reach converged estimates. Here, quantum mechanics/molecular mechanics (QM/
MM) molecular dynamics (MD) simulations play an important role, providing QM accuracy for the region of interest at a decreased computational cost. However, QM/MM simulations are still too expensive to study large systems on longer time scales. Recently, machine learning (ML) models have been proposed to replace the QM description. The main limitation of these models lies in the accurate description of long-range interactions present in condensed-phase systems. To overcome this issue, a recent workflow has been introduced combining a semi-empirical method (i.e. density functional tight binding (DFTB)) and a high-dimensional neural network potential (HDNNP) in a $\Delta$-learning scheme. This approach has been shown to be capable of correctly incorporating long-range interactions within a cutoff of 1.4 nm. One of the promising alternative approaches to efficiently take long-range effects into account is the development of graph convolutional neural networks (GCNN) for the prediction of the potential-energy surface. In this work, we investigate the use of GCNN models -- with and without a $\Delta$-learning scheme -- for (QM)ML/MM MD simulations. We show that the $\Delta$-learning approach using a GCNN and DFTB and as baseline achieves competitive performance on our benchmarking set of solutes and chemical reactions in water. The method is additionally validated by performing prospective (QM)ML/MM MD simulations of retinoic acid in water and S-adenoslymethioniat interacting with cytosine in water. The results indicate that the $\Delta$-learning GCNN model is a valuable alternative for (QM)ML/MM MD simulations of condensed-phase systems.

\section{Introduction}

A key goal of computational chemistry is the molecular level understanding of chemical reactions in solution and enzymes. 
For this, the free-energy change (rather than the change in potential energy) during a reaction process is the central property. To calculate free-energy differences, molecular dynamics (MD) simulations from tens of picoseconds to hundreds of nanoseconds are typically required to obtain sufficiently converged results. While classical force fields can be used for MD simulations of condensed-phase systems over long time scales,\cite{Riniker2018Fixed-ChargeOverview, Nerenberg2018NewSimulations} higher-level quantum-mechanical (QM) methods are required for an accurate description of molecular interactions and chemical reactions. Unfortunately, QM calculations are much more computationally intensive, limiting the accessible time and spatial scales.

In order to overcome this bottleneck, the combined QM and molecular mechanical (QM/MM) approach provides a QM description of the region of interest (QM zone) coupled with a realistic modelling of the long-range interactions of the surrounding condensed-phase system (MM zone).\cite{Warshel1976TheoreticalLysozyme, Mulholland2000AbInvolved, Senn2009QM/MMSystems, Groenhof2013IntroductionSimulations} The interaction between the zones is then calculated based either on mechanical constraints (i.e., "mechanical embedding" scheme) or electronic perturbations (i.e. "electrostatic embedding" scheme). Generally, the electrostatic embedding scheme has been shown to be more accurate and it is currently the gold standard for QM/MM simulations\cite{Senn2009QM/MMSystems, Lin2007QM/MM:Here, Panosetti2020LearningRegression, Brunk2015MixedStates, Chung2015TheApplications}. 
In the QM/MM scheme, the QM zone requires electronic structure calculations at each time step and is thus the computational bottleneck. While the computational costs of QM/MM MD simulations are reduced compared to full \textit{ab initio} simulations, the accessible time and spatial scales are still not sufficient for most free-energy calculations. This issue can be partially circumvented by using semi-empirical methods to describe the QM zone\cite{Brunk2015MixedStates, Christensen2016SemiempiricalApplications, Stewart2013OptimizationParameters}. However, this reduces not only the computational cost but also the achievable accuracy. An alternative is to use machine-learned (ML) potentials to describe the QM zone. 

In recent years, there have been major advances in the development of ML models trained to reproduce the potential-energy surface (PES) of chemical systems\cite{Rupp2012, Behler2011, Hansen2013, Lorenz2004, Behler2014, Smith2018, Behler2016,  Smith2017, Grisafi2021Multi-scaleProperties,  Schutt2017SchNet:Interactions, Unke2019PhysNet:Charges, KovacsLinearRMSE, Zaverkin2020, Ko2021ATransfer, Li2015, Anderson2019Cormorant:Networks}. For small to medium sized compounds in the gas phase or in periodic materials, state-of-the-art ML models have been shown to achieve chemical accuracy, both for the predicted energies and forces \cite{Smith2017, Schutt2017SchNet:Interactions, Unke2019PhysNet:Charges, KovacsLinearRMSE, Grisafi2021Multi-scaleProperties, Xie2018CrystalProperties}. 
However, condensed-phase systems pose an additional challenge to ML models due to the typically large number of involved atoms, element types, and rotatable bonds without any exploitable symmetries, as well as important long-range interactions and non-local charge transfer. In particular, the long-range interactions and the non-local charge transfer currently limit the established ML methods, as these often rely on local descriptors and are thus unable to take global changes in the electronic structure into account\cite{Grisafi2019IncorporatingLearning, Ko2021ATransfer, Unke2021SpookyNet:Effects}.
There are two main approaches to overcome this locality dependence of the descriptors used in current ML models. Either non-local information transfer is incorporated directly in the ML models\cite{Grisafi2019IncorporatingLearning, Chen2019GraphCrystals, Ko2021ATransfer, Unke2021SpookyNet:Effects} or ML methods are combined with a fast or semi-empirical QM method with an explicit treatment of long-range interactions in a $\Delta$-learning scheme \cite{Ramakrishnan2015BigApproach,Boselt2021MachineSystems, Shen2018MolecularNetworks, Zeng2021DevelopmentSolution}. The $\Delta$-learning approach has already been shown to be promising for (QM)ML/MM MD simulations of condensed-phase systems\cite{Boselt2021MachineSystems, Shen2018MolecularNetworks, Zeng2021DevelopmentSolution}. 

Next to the descriptor-based approaches, there has also been a lot of development on message passing approaches trained to reproduce the PES of QM systems\cite{Schutt2017SchNet:Interactions, Unke2019PhysNet:Charges, Schutt2021EquivariantSpectra, Klicpera2020DirectionalGraphs,Satorras2021EnNetworks}. These graph networks typically use dense layers of neural networks as non-linear functions for the message passing convolutions and are thus known as graph-convolutional neural networks (GCNNs).
One of the main advantages of GCNNs compared to descriptor-based ML models is that no specialized descriptors have to be developed for the chemical systems. Instead, this is achieved directly through the graph-convolutional layers. A further advantage is that through iterative message passing operations more distant information is taken into account and thus the local dependence of descriptor based models can be (partially) avoided. However, each consecutive convolution substantially increases the model size. Thus, in practice long-range contributions are terminated at a certain cutoff and global changes in the electronic structure are not considered.
To address this issue, the total energy can be separated into a short-range and a long-range electrostatic term, for which a GCNN is used to predict atomic charges\cite{Unke2019PhysNet:Charges, Unke2021SpookyNet:Effects}. Further, long-range charge transfer and global changes in the electronic structure may be taken into account by including a global state or a self-attention mechanism in the message passing operation\cite{Chen2019GraphCrystals, Unke2021SpookyNet:Effects, Unke2019PhysNet:Charges, Vaswani2017AttentionNeed}.

In this work, we assess the applicability of GCNNs to reproduce the PES of condensed-phase systems and their use in (QM)ML/MM MD simulations. For this, we validate a GCNN model with and without a $\Delta$-learning scheme on different molecular systems in water and compare it with the previously developed high-dimensional neural network potentials (HDNNPs) ,\cite{Behler2011Atom-centeredPotentials, Boselt2021MachineSystems} which uses the same $\Delta$-learning scheme with DFTB \cite{Elstner2006TheSystems, Hourahine2020DFTB+Simulations} as baseline method. In the Theory section, we briefly describe the relevant concepts behind the QM/MM approach as well as the GCNNs. In the Methods section, we describe the setup used for the QM/MM simulations, the implemented GCNN architecture as well as the training setup. In the Results section, we evaluate different GCNN models, including different global information transfer schemes, with different (QM)ML/MM and training setups. Finally, we compare the resulting GCNN model with the previous HDNNP model\cite{Boselt2021MachineSystems}.

\section{Theory}
\subsection{QM/MM Scheme}
QM/MM is a multi-scale approach, which incorporates a QM zone within a larger MM zone describing the condensed-phase system\cite{Warshel1976TheoreticalLysozyme, Mulholland2000AbInvolved, Senn2009QM/MMSystems, Groenhof2013IntroductionSimulations}. This enables the accurate calculation of the region of interest coupled with a realistic modelling of long-range interactions with the surrounding environment. The main challenge here is how to describe the interaction between these two different zones. In order to calculate the total energy ($E_{QM/MM}(\vec{R})$) of the combined QM/MM system, an additive or subtractive scheme can be chosen. 
In the more prominent additive scheme, the total energy ($E_{QM/MM}(\vec{R})$) is described as sum of the energy of the QM- ($E_{QM}(\vec{R}_{QM})$) and MM-subsystems ($E_{MM}(\vec{R}_{MM})$) plus the electrostatic ($E^{el}_{QM-MM}(\vec{R})$) and short-range van der Waals interactions ($E^{vdW}_{QM-MM}(\vec{R})$) between the two subsystems. 
\begin{equation}
  E_{QM/MM}(\vec{R}) = E_{QM}(\vec{R_{QM}}) + E_{MM}(\vec{R}_{MM}) + E^{el}_{QM-MM}(\vec{R}) + E^{vdW}_{QM-MM}(\vec{R}) \label{eqn:additive}
\end{equation}
Note the distinction between $\vec{R}$ referring to all nuclei in the system, and $\vec{R}_{QM}$ and $\vec{R}_{MM}$ referring to the nuclei of the QM and MM zone respectively.
In the additive scheme, the interaction terms between the QM and MM zones are described either via a mechanical or an electrostatic embedding scheme\cite{Lin2007QM/MM:Here}, where the latter scheme has been shown to be more accurate\cite{Senn2009QM/MMSystems}.
For this, two Hamiltonians are introduced in the QM calculation. In atomic units, $\hat{H}^{el}_{QM-MM}$ is given as,
\begin{equation}
 \hat{H}^{el}_{QM-MM} = -\sum^{N_{MM}}_i\sum^{N_{el}}_j\frac{q_i}{|\vec{R}_{MM,i}-\vec{r}_j|}  
 +\sum^{N_{QM}}_i\sum^{N_{MM}}_j\frac{Z_iq_j}{|\vec{R}_{QM,i}-\vec{R}_{MM,j}|} \label{eqn:Hel}
\end{equation}
where $\hat{H}^{vdW}_{QM-MM}$ is treated classically and is given as,
\begin{equation}
  \hat{H}^{vdW}_{QM-MM} = E^{vdW}_{QM-MM}(\vec{R}) = \sum^{N_{QM}}_i\sum^{N_{MM}}_j 4\epsilon_{ij}\left(\left(\frac{\sigma_{ij}}{|\vec{R}_i-\vec{R}_j|}\right)^{12} - 
  \left(\frac{\sigma_{ij}}{|\vec{R}_i-\vec{R}_j|}\right)^{6}\right) \label{eqn:HvdW}, 
\end{equation}
where $q_i$ is the partial charge of MM atom $i$, and $\epsilon_{ij}$ and $\sigma_{ij}$ are fitted parameters. This means that the QM subsystem is directly influenced by the MM partial charges, while the MM subsystem ``feels'' a force from the perturbed QM subsystem. Therefore, Eq.~(\ref{eqn:additive}) becomes in the electrostatic embedding scheme,
\begin{equation}
  E_{QM/MM}(\vec{R}) = \frac{\braket{\psi(\vec{r})|(\hat{H}_{QM}+\hat{H}^{el}_{QM-MM})\psi(\vec{r})}}{\braket{\psi(\vec{r})|\psi(\vec{r})}} + E_{MM}(\vec{R}_{MM}) +  E^{vdW}_{QM-MM}(\vec{R}) \label{eqn:electrostatic}
\end{equation}
Note that only MM particles within a given cutoff radius $R_c$ of the QM zone are included in the summations in Eqs.~(\ref{eqn:Hel}) and (\ref{eqn:HvdW}). Typically, a relatively large cutoff radius ($R_c$) of around 1.4~nm is required to achieve sufficiently converged results\cite{Panosetti2020LearningRegression, Boselt2021MachineSystems}. The arising issue due to the non-continuous PES at the cutoff can be partially resolved using adaptive resolution schemes\cite{Bulo2009TowardSimulations}. 

\subsection{Graph-Convolutional Neural Networks}
In our description of graph-convolutional neural networks (GCNNs) or message passing neural networks, we follow the notations used by Refs.~\citenum{Gilmer2017NeuralChemistry, Satorras2021EnNetworks}. Here, GCNNs are permutation-invariant ML models that operate on graph structured data. In the present work, atoms are represented by
nodes $\nu$ and interactions between atoms as edges $e$, whereby only interaction up to a certain cutoff $R_\text{edge}$ are considered as edges. Note that we will refer to two different cutoffs throughout the manuscript. $R_c$ is the cutoff used in the QM/MM scheme, and the summations in  Eqs.~(\ref{eqn:Hel}) and (\ref{eqn:HvdW}) run over all partial charges within $R_c$. $R_\text{edge}$, on the other hand, is the cutoff used for the edge definition in the GCNNs. 
GCNNs contain consecutive graph-convolutional layers, where each layer consists of an edge or message update operation (Eq.~(\ref{eqn:message})), an aggregation operation (Eq.~(\ref{eqn:aggregation})), and a node update operation (Eq.~(\ref{eqn:node})). Considering a graph $G = (V,E)$ with nodes $\nu_i \in V$ and edges $e_{ij} \in E$, message passing can be defined as,
\begin{equation}
  m_{ij} = \phi_e(h^l_i,h^l_j,a_{ij})\label{eqn:message}
\end{equation}
\begin{equation}
  m_i = \sum_{j\in N(i)}m_{ij}\label{eqn:aggregation}
\end{equation}
\begin{equation}
  h^{l+1}_i = \phi_h(h^l_i, m_i)\label{eqn:node}.
\end{equation}
Here, the superscript $l$ denotes the current layer, $h^l_i \in \textbf{R}^n$ describes the hidden-feature vector of node $\nu_i$ at layer $l$, $a_{ij} \in \textbf{R}^n$ describes the edge feature of edge $e_{ij}$ between nodes $i$ and $j$, $N(i)$ denotes the set of neighbors of node $i$, and $\phi_e$ and $\phi_h$ describe update functions. 
Different GCNN models commonly differ by their used features, update functions ($\phi_e$ and $\phi_h$), and aggregation functions.\cite{Gilmer2017NeuralChemistry, Satorras2021EnNetworks}. The update functions are most commonly approximated by multilayer perceptrons. Note that we use a summation as aggregation function (Eq. ~(\ref{eqn:aggregation})), but other aggregation functions such as min or max functions have also been investigated\cite{Corso2020PrincipalNets}.

\section{Methods}
\subsection{Systems}
To allow a direct comparison, the same systems as in Ref.~\citenum{Boselt2021MachineSystems} were investigated. For the validation of the different GCNN models and the training setups, we used two single-solute systems with different MM cutoff radii ($R_c$): (i) benzene in water, and (ii) uracil in water. 
For the comparison to the previous HDNNP model,\cite{Boselt2021MachineSystems} we used the largest single-solute system (retionic acid in water), and two chemical reactions in water (constrained close to the transition state): (i) the second-order nucleophilic substitution ($S_N2$) reaction of \ce{CH3Cl} with \ce{Cl-}, and (ii) the reaction of S-adenosylmethionate (SAM) with cytosine. We investigated the accuracy of the different models in training/validation/test splits as well as the performance in a prospective (QM)ML/MM MD simulation.
The data sets are freely available on \url{https://www.research-collection.ethz.ch/handle/20.500.11850/512374}.

\subsection{General Computational Details}
All QM/MM and (QM)ML/MM MD simulations were performed using the GROMOS software package \cite{Schmid2012ArchitectureSimulation, Meier2012InterfacingPackages} interfaced to DFTB+/19.2\cite{Elstner2006TheSystems, Hourahine2020DFTB+Simulations} and ORCA/4.2.0\cite{Neese2018Software4.0}. All structures used in the training, validation and test sets were taken from Ref.~\citenum{Boselt2021MachineSystems} (available on \url{https://www.research-collection.ethz.ch/handle/20.500.11850/512374}). These come from QM/MM MD trajectories, where the first 70\% of the 10'000 frames were considered to be the training set, the following 20\% constitute the validation set, and the last 10\% are taken as test set. The computational details for the (QM)ML/MM simulations are provided in Ref.~\citenum{Boselt2021MachineSystems}. Note that a MM cutoff radius of $R_c = 0.6$ nm was used for benzene in water , while $R_c = 1.4$~nm was used for all other systems. 
The GCNNs were implemented using Tensorflow/keras \cite{Tensorflow2015-whitepaper}. The models were trained in Python and then exported to the C++ GROMOS code as described in Ref.~\citenum{Boselt2021MachineSystems} for the HDNNPs. 

For the validation of the different GCNN models, we compare both the full-QM learning task and the $\Delta$-learning scheme. For the full-QM learning task, we use the GCNN models to directly predict the DFT \cite{Neese2018Software4.0} energies and forces, whereas DFTB \cite{Elstner2006TheSystems, Hourahine2020DFTB+Simulations} is used as baseline method in the $\Delta$-learning scheme and the GCNN models predicts the difference between the DFT and the DFTB properties. After the initial validation and comparison of the two learning tasks, we continue only with the $\Delta$-learning approach throughout the remainder of the study.

\subsection{(QM)ML/MM MD Simulations}
For the retionic acid in water and the SAM/cytosine transition state in water, we performed (QM)ML/MM MD simulations using the trained GCNN models with the $\Delta$-learning setup. To ensure comparability, we used the same MD and DFTB settings as in Ref.~\citenum{Boselt2021MachineSystems}. The time step was set to 0.5~fs, the temperature to $T$~=~298~K, and the pressure to 1~bar. For the SAM/cytosine system, we set the force constant for the position restraints to 2000~kJ~mol$^{-1}$~nm$^{-2}$. All point charges within the cutoff radius $R_c$~=~1.4~nm were included in an electrostatic embedding scheme in the QM(ML) computation. Selected solvent atoms beyond the cutoff were included to avoid bond-breaking and creation of artificial charges. Long-range electrostatic interactions beyond $R_c$ were included using a reaction-field method \cite{Tironi1995ASimulations}. Note that the reaction field acts only on the MM particles. As starting coordinates for the prospective simulations, we used the last snapshot from the test set (which originated from the initial QM/MM MD trajectory). The simulations were performed for 200'000 steps for retionic acid in water and for 50'000 steps for the SAM/cytosine transition state in water. 

\subsection{GCNN Architecture}
\begin{figure}[H]
  \centering
  \includegraphics[width=0.8\textwidth]{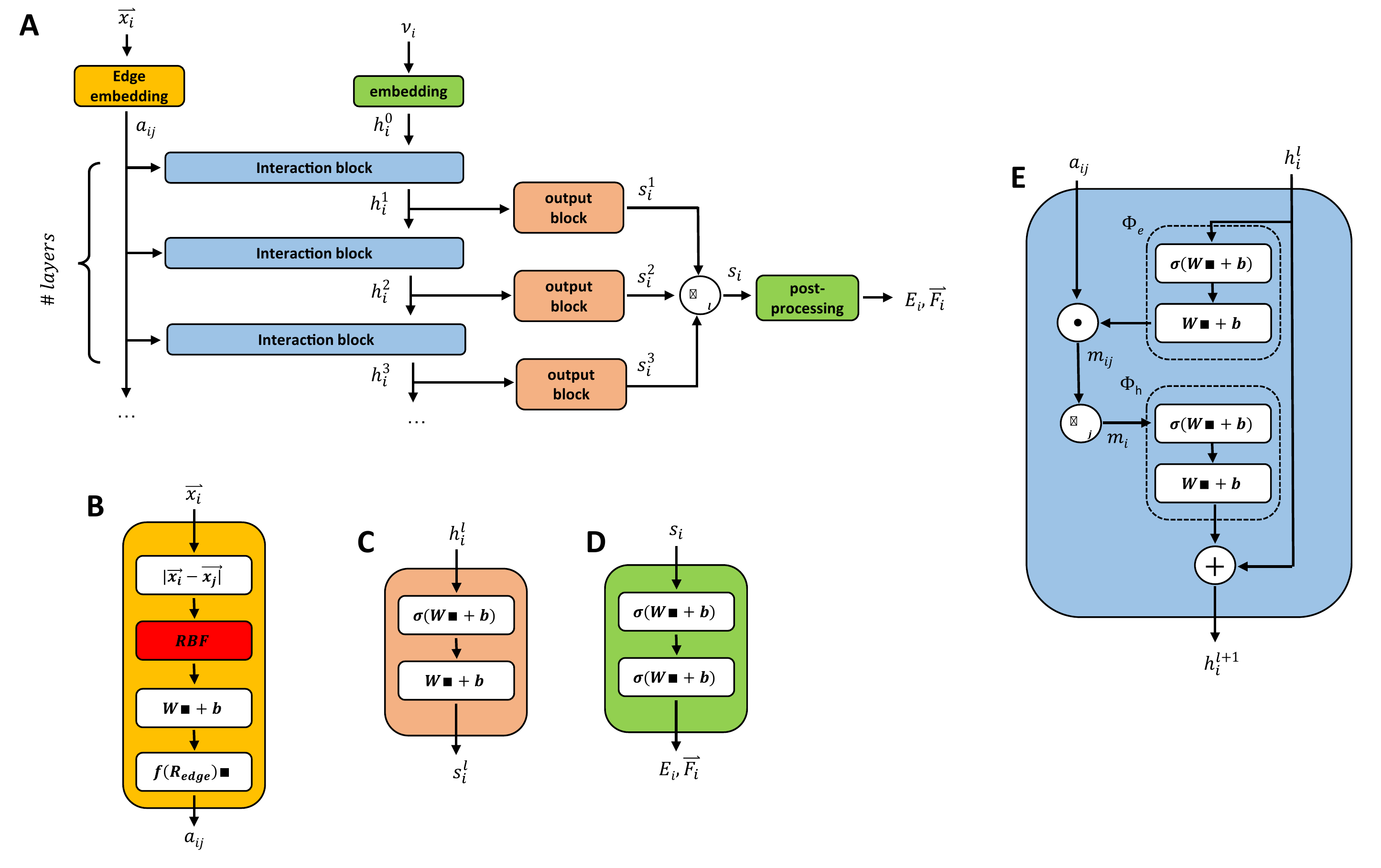}
  \caption{Overview of the employed GCNN models with the full architecture (\textbf{A}), the cfconv block (\textbf{B}), the output block (\textbf{C}), the post-processing block (\textbf{D}), and the interaction block (\textbf{E}). For all linear layers $Wx + b$, we use $n_\text{f}$. For all activation functions $\sigma()$, we use a Swish activation function.}
  \label{fgr:GNN_arch}
\end{figure}

 The basic building blocks of the GCNN are so-called dense layers. They take an input vector $x \in R^{n_{in}}$ and return an output vector $y \in R^{n_{out}}$ according to the transformation,
\begin{equation}
  y = Wx + b \label{eqn:dense}.
\end{equation}
Here, $W \in R^{n_{in}\times n_{out}}$ and $b \in R^{n_{out}}$ are learnable parameters. In order to model arbitrary non-linear relationships, at least two dense layers need to be stacked and combined with a (non-linear) activation function $\sigma$. Here, we use a generalized SiLU (Sigmoid Linear Unit) activation known as Swish activation function \cite{DBLP:journals/corr/abs-1710-05941}, which is given as $\sigma(x) = x  \cdot \mathrm{sigmoid}(x)$ and has been used successfully in recently published GCNNs for molecular systems.\cite{Klicpera2020DirectionalGraphs, Unke2021SpookyNet:Effects, Satorras2021EnNetworks} 
The inputs to the GCNN ($\nu_i$) are the nuclear charges $Z_i \in \mathbb{N}$ and positions $\vec{r}_i \in R^3$.
We evaluate four different GCNN architectures with different global information transfer schemes. An overview of the basic network architecture is given in Figure \ref{fgr:GNN_arch}A. In the following sections, this architecture is referred to as GCNN model. 

\textit{Embedding block}. An embedding is a mapping from a discrete object to a vector of real numbers. Here, the atomic numbers are mapped to embeddings $e_Z \in R^F$, where the entries of $e_Z$ are learnable parameters and $n_f$ denotes the number of features. Note that the number of features is kept constant throughout the network. The embedding vector is then used to initialize the atomic feature vector $h^0_i$.

\textit{Edge embedding block}. A continuous filter convolution block is used to generate the edge representations $a_{ij}$ (Figure \ref{fgr:GNN_arch}B). First, all edges are expressed as Euclidean distances. These are subsequently transformed to linear combinations of rationally-invariant filters (RBFs) of radial basis functions $sin(\frac{n\pi}{R_\text{edge}}||\vec{r}_{ij}||/||\vec{r}_{ij}||$ as proposed by \citeauthor{Klicpera2020DirectionalGraphs}\cite{Klicpera2020DirectionalGraphs}. Additionally, we apply a cosine cutoff to the filters \cite{Behler2011Atom-centeredPotentials}, which ensures continuous behavior when an atom enters or leaves the cutoff sphere.

\textit{Interaction block}. The interaction blocks calculate the message passing operation as defined in Eqs. (\ref{eqn:message})-(\ref{eqn:node}) (Figure \ref{fgr:GNN_arch}E), generating the atomic feature vectors $h^l_i$ at layers $l$. 

\textit{Output block}. The atomic feature vectors $h^l_i$ at each layer $l$ are passed through an output block, consisting of two stacked dense layers combined with an activation function ($\sigma$) (Figure \ref{fgr:GNN_arch}C). The output vectors $s^l_i$ of each layer $l$ are then summed and passed through a post-processing block, consisting of two dense layers, each combined with an activation function (Figure \ref{fgr:GNN_arch}D). The output of the post-processing block are the atomic energies. Finally, the total energy is calculated as sum over all atomic energies. The forces are subsequently obtained as derivatives of the total energy with respect to the Cartesian coordinates of the atoms. For this, we use the reverse mode automatic differentiation implemented in Tensorflow. 

\begin{figure}[H]
  \centering
  \includegraphics[width=0.8\textwidth]{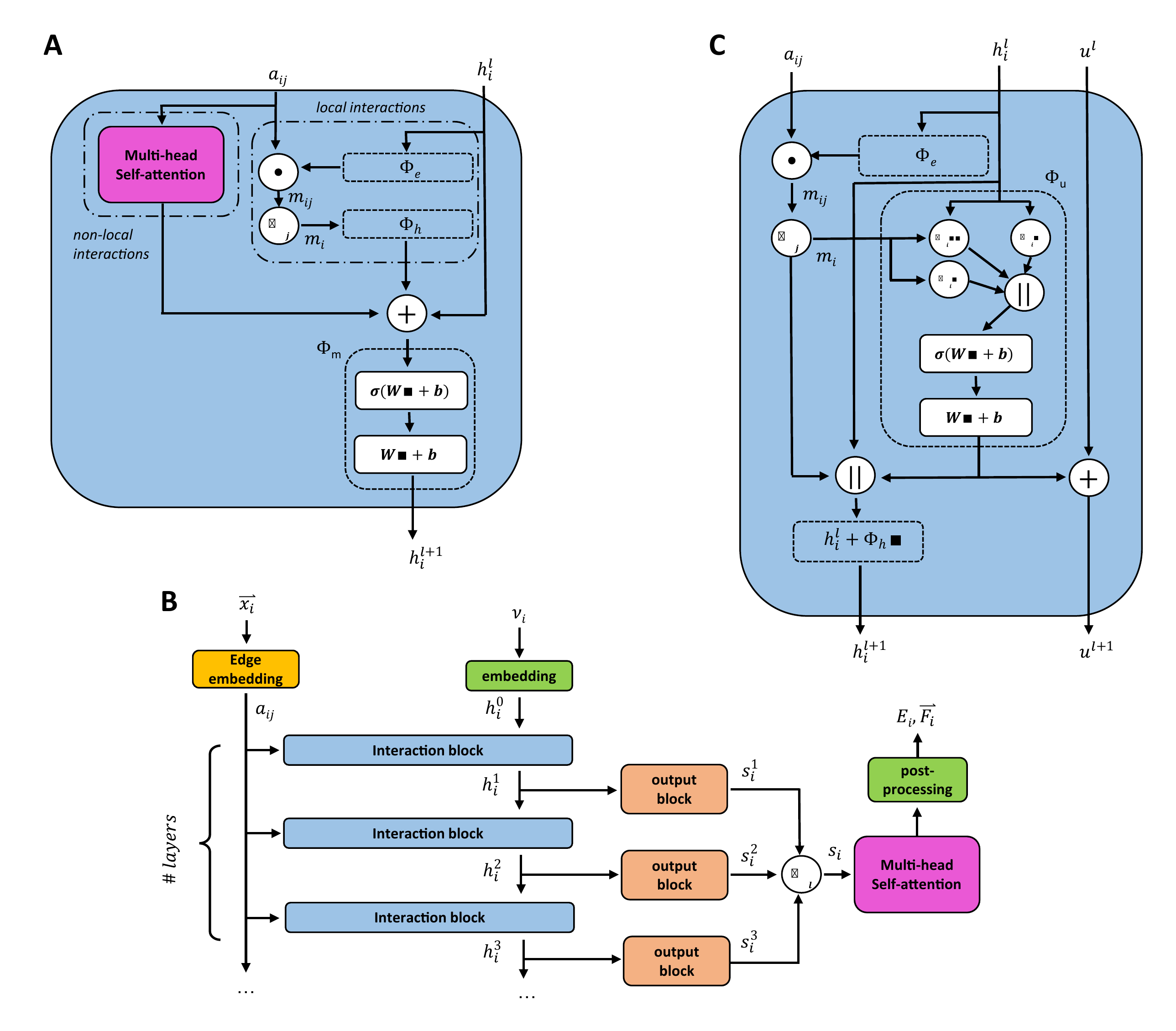}
  \caption{Changes in the architecture for the GCNN models with different global information transfer schemes. (\textbf{A}): Interaction block for the GCNN model with multi-head self-attention layer in the interaction block. (\textbf{B}): Full architecture for the GCNN model with multi-head self-attention layer prior to the post-processing block. (\textbf{C}): Interaction block for the GCNN model with global state $u^l$. }
  \label{fgr:GNN_global_arch}
\end{figure}

Unke \textit{et al.}\cite{Unke2021SpookyNet:Effects} introduced a GCNN model, which takes into account non-local effects and charge transfer. They achieved this by introducing non-local interactions using a self-attention layer \cite{Vaswani2017AttentionNeed, CordonnierMulti-HeadConcatenate}. An attention layer maps a matrix $X \in R^{T \times n_{in}}$ of query ($Q$) tokens $T$ to $n_{out}$ dimensions  using a matrix $Y \in R^{T'\times n_{in}}$ of key ($K$) tokens $T'$ as follows,
\begin{equation}
  \text{Attention}(Q,K,V) = \sigma \left( \frac{QK^T}{\sqrt(d_k)}\right)V \label{eqn:attention}
\end{equation}
\begin{equation}
Q = XW_Q, K = YW_k, V=YW_V \label{eqn:attention2}.
\end{equation}
The layer is parametrized by a query matrix $W_Q \ in R^{n_{in}\times n_k}$, a key matrix $W_K \ in R^{n_{in}\times n_k}$, and a value matrix $W_V \ in R^{n_{in}\times n_{out}}$. A self-attention layer uses attention on the same sequence $(X=Y)$.
In this work, we use a multi-head self-attention mechanism as described in Ref.~\citenum{CordonnierMulti-HeadConcatenate}. Here, the attention is calculated for multiple heads using a reduced attention $n_{out} = n_k < n_{in}$ dimensionality. The multiple attentions are then concatenated and transformed as, 
\begin{equation}
\text{MultiAttention} = \left\{\text{Attention}_0 | \text{Attention}_1| ...| \text{Attention}_{N_\text{heads}}\right\} W_\text{multi}, \label{eqn:attention3}.
\end{equation}
where $W_\text{multi} \in R^{N_\text{heads}n_{out}\times n_\text{multi}}$ is used to parametrize the layer. 

We introduce the multi-head self-attention layer at two different stages in our basic GCNN architecture: (i) within the interaction block (Figure \ref{fgr:GNN_global_arch}A) as described in Ref.~\citenum{Unke2021SpookyNet:Effects}, and (ii) after at the end of the model (Figure \ref{fgr:GNN_global_arch}B). This way we can investigate the effect of repeating the global information transfer within each message passing update as compared to a single transfer step using the final atom features. In the following sections, these models are referred to as ``interaction GCNN'' and ``attention GCNN''.

A different GCNN architecture, which considers global information transfer, has been reported by Chen \textit{et al.}\cite{Chen2019GraphCrystals} Here, the authors introduce a global state $u^l$ into the message passing operation. We adapted this idea by changing the message update step (Eq.~(\ref{eqn:node})) to,
\begin{equation}
u^{l+1} = u^l + \phi_u (\sum_i h^l_i, \sum_i m_i, \sum_i h^l_i m_i) \label{eqn:global}
\end{equation}
\begin{equation}
h^{l+1} = h^l + \phi_h (h^l_i, m_i, u^{l+1}) \label{eqn:global2},
\end{equation}
where $u^0 \in R^F$ is initialized as zero. The updated interaction block is shown in Figure \ref{fgr:GNN_global_arch}C. In the following sections, we refer to this architecture as ``global GCNN''.

\subsection{Training Setup}
We implemented the four different GCNN model architectures using Tensorflow/keras. The models were trained using the Adam optimizer \cite{kingma2017adam} with an exponentially decaying learning rate ([initial learning-rate, decay steps, decay rate] = [1$e^{-3}$, 5$e^3$, 0.96]). The Adam optimizer is one of the most well-established methods for the training of neural network models. It is based on a stochastic gradient descent optimizer that uses an adaptive estimation of first and second-order moments. 
The models were trained for up to 2'000 epochs (or until convergence) using a batch sampling of 2-32, depending on the memory requirements of the models and systems. Note that the models were trained using Tensorflow 2.7.0, but the models were saved in Tensorflow 1.15 for the integration with the GROMOS C++ code (as described in Ref.~\citenum{Boselt2021MachineSystems} for the HDNNP's).

As loss function, we used the weighted mean-squared-error (MSE) for the energies and forces,
\begin{equation}
L = \frac{1}{N}\sum_i^N(E_i-\Tilde{E}_i)^2 
+ \frac{\omega_0}{3N_{QM}}\sum_i^{N_{QM}}\sum_\alpha^3(F_{i\alpha}-\Tilde{F}_{i\alpha})^2
+ \frac{\omega_0}{3N_{MM}}\sum_i^{N_{MM}}\sum_\alpha^3(F_{i\alpha}-\Tilde{F}_{i\alpha})^2\label{eqn:loss}
\end{equation}
where $N_{QM}$ is the number of QM particles, $N_{MM}$ is the number of MM particles, and $\omega_0$ and $\omega_1$ are weight parameters for the gradient contributions. We monitored the loss during the training process and recovered the model with the lowest loss on the validation set after training.

\section{Results and Discussion}
In order to evaluate the different model architectures and parametrizations as well as the training setup, the two simplest test systems from Ref.~\citenum{Boselt2021MachineSystems}, i.e., benzene in water and uracil in water, were used with the same training/validation/test split as in the original publication. The use case here is that the training set is generated from a short initial MD simulation, from which the energies and forces of the subsequent MD steps can be predicted. In the first step, we evaluated the full-QM learning task as well as a $\Delta-$learning scheme with DFTB \cite{Elstner2006TheSystems, Hourahine2020DFTB+Simulations} as baseline for the different GCNN architectures. The $\Delta-$learning scheme simplifies the learning task, and we adopted the same approach as used in Ref.~\citenum{Boselt2021MachineSystems} for the HDNNPs.  
For the basic setup of our GCNN model, 128 features per dense layer ($n_f$), five interaction blocks, and $r_{cut}$~=~0.5~nm were used. For the multi-head self-attention models, we used four heads with $n_k=32$. Finally, we compare the best GCNN model with the previous HDNNP model in Ref.~\citenum{Boselt2021MachineSystems} for all five test systems: (i) benzene in water, (ii) uracil in water, (iii) retionic acid in water, (iv) (close to) transition state of the $S_N2$ reaction of \ce{CH3Cl} with \ce{Cl-} in water, and (v) (close to) transition state of SAM with cytosine in water.
Note that for the parametrization and evaluation of the GCNN models and training procedure, we used solely the training and validation sets. The test sets were only taken for the comparison between the final GCNN model with the previous HDNNP model.

\subsection{Model Architecture}
In a first step, we compared the different model architectures in order to evaluate how the global transfer schemes influence the inclusion of the long-range information directly into the model. For this, four different GCNN models were trained: (a) a basic GCNN model, (b) a GCNN model including multi-head self-attention at the post-processing stage (labeled as ``attention GCNN''), (c) a GCNN model including multi-head self-attention in the interaction layers (labeled as ``interaction GCNN''), and (d) a GCNN model including a global state (labeled as ``global GCNN''). The two test systems consist of benzene (apolar molecule) in water with a short MM cutoff radius of $R_c = 0.6$~nm, and uracil (polar molecule) in water with a long $R_c = 1.4$~nm.

\begin{table}[H]
  \caption{Mean absolute error (MAE) on the validation set (2'000 frames) for GCNN models containing global information for the test systems benzene in water and uracil in water. For each property, the model with the lowest MAE is marked in bold.}
  \label{tbl:model_type}
 \centering
 \small
  \begin{tabular*}{\textwidth}{l | rrr | rrr}
    \hline
    \multicolumn{7}{l}{\textbf{Benzene}} \\ \hline
    & \multicolumn{3}{c|}{\textbf{full QM}} & \multicolumn{3}{c}{\textbf{$\Delta-$learning}} \\
    GCNN & $E $  & $F_{QM} $ & $F_{MM} $ & $E $  & $F_{QM} $& $F_{MM} $ \\
    model & kJ~mol$^{-1}$  & kJ~mol$^{-1}$~nm$^{-1}$ & kJ~mol$^{-1}$~nm$^{-1}$ & kJ~mol$^{-1}$  & kJ~mol$^{-1}$~nm$^{-1}$ & kJ~mol$^{-1}$~nm$^{-1}$\\
    \hline
    basic & 2.7 & \textbf{48.7} & \textbf{8.3} & \textbf{1.0} & 22.9 & \textbf{3.5}\\
    global & \textbf{2.4} & 54.1 & 8.9 & \textbf{1.0} & \textbf{22.7} & 3.8\\
    interaction & 2.7 & 52.8 & 9.7 & 3.9 & 36.2 & 7.3\\
    attention & 2.7 & 61.7 & 10.0 & 1.2 & 23.7 & 4.2\\
    \hline
    \hline
    \multicolumn{7}{l}{\textbf{Uracil}} \\ \hline
    & \multicolumn{3}{c|}{\textbf{full QM}} & \multicolumn{3}{c}{\textbf{$\Delta-$learning}} \\
    GCNN & $E $  & $F_{QM} $ & $F_{MM} $ & $E $  & $F_{QM} $& $F_{MM} $ \\
    model & kJ~mol$^{-1}$  & kJ~mol$^{-1}$~nm$^{-1}$ & kJ~mol$^{-1}$~nm$^{-1}$ & kJ~mol$^{-1}$  & kJ~mol$^{-1}$~nm$^{-1}$ & kJ~mol$^{-1}$~nm$^{-1}$\\
    \hline
    basic & \textbf{7.3} & 135.2 & \textbf{5.8} & \textbf{2.2} & 60.2 & \textbf{1.3}\\
    global & 8.1 & \textbf{129.7} & 6.4 & 2.3 & 59.8 & 1.7\\
    interaction & 8.0 & 138.4 & 4.4 & 2.6 & \textbf{55.5} & 1.6\\
    attention & 7.4 &  134.9 & 7.3 & 2.5 & \textbf{55.5} & 1.4\\
    \hline
  \end{tabular*}
\end{table}

\begin{figure}[H]
  \includegraphics[width=\textwidth]{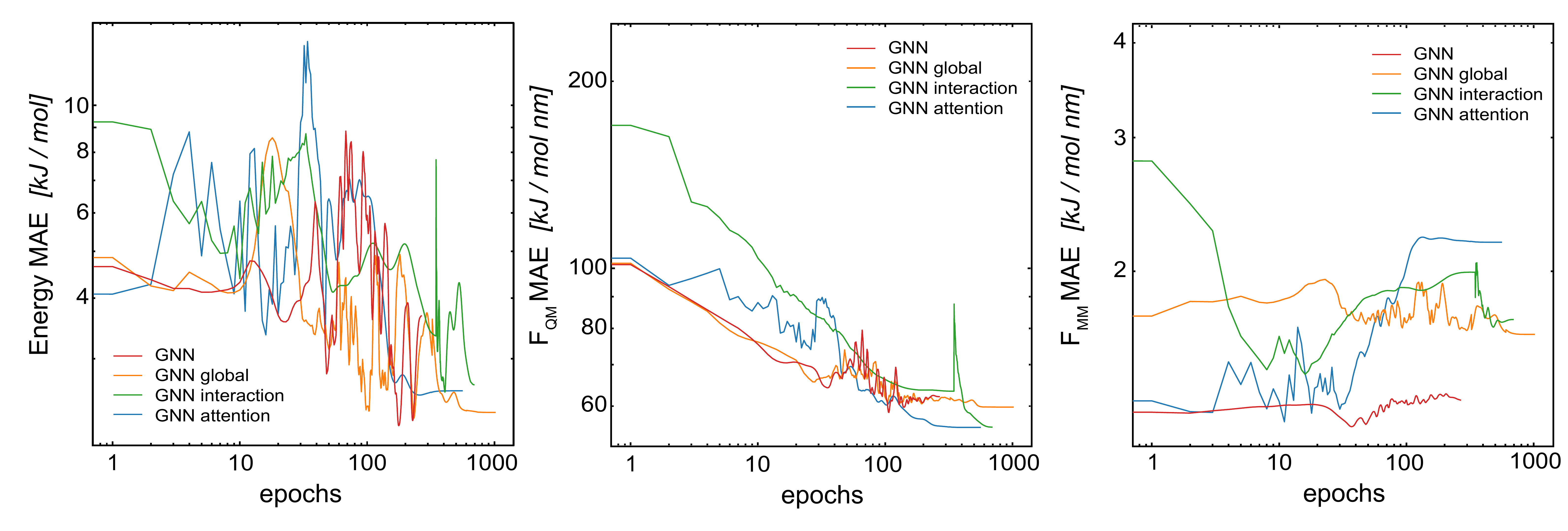}
  \caption{Mean absolute error (MAE) on the validation set as a function of the training epochs (learning curves) for the $\Delta$-learning GCNN models containing global information for the test system uracil in water. (\textbf{Left}): MAE of the energies in the QM zone. (\textbf{Middle}): MAE of the forces on the QM particles. (\textbf{Right}): MAE of the forces on the MM particles from the QM zone.}
  \label{fgr:model_type}
\end{figure}

Table \ref{tbl:model_type} shows the mean absolute error (MAE) of the different model architectures for the validation sets of benzene in water and uracil in water. When the models were trained on the full QM energies and forces, we observe for benzene that all models achieve a similar accuracy on the energies, while the basic GCNN outperforms the other architectures for the forces ($F_{QM}$ and $F_{MM}$). When using the $\Delta$-learning scheme, the accuracy is generally improved, with similar performances of the different architectures (Table \ref{tbl:model_type}). Only the interaction GCNN model has a significantly higher MAE for both the energies and the forces.
For uracil, we observe similar trends as for benzene (Table \ref{tbl:model_type}). Again, no large differences are observed between the models for both the ``full QM'' as well as the $\Delta$-learning setup (Figure \ref{fgr:model_type}). In contrast to benzene, also the interaction GCNN models achieve similar accuracy compared to the other models.

As observed for the HDNNPs in Ref.~\citenum{Boselt2021MachineSystems}, the use of a $\Delta$-learning scheme with DFTB as baseline method reduces the prediction error by about two folds. This is also the case for the GCNN models containing global information. 
This indicates that the $\Delta$-learning approach is more suited to incorporate the long-range interactions. A reason for this could be that while the GCNNs containing global information can in theory learn long-range interactions, the large number of MM atoms present lead to an exponential increase in possible system configurations. This in turn requires a huge number of training data points in order to accurately capture the long-range interactions. If the size of the training set is limited to a practically useful number (as done here), the training set is evidently not large enough to achieve the desired accuracy.

In conclusion, the $\Delta$-learning scheme leads to a clear and consistent performance improvement, whereas no large differences are observed between the four different GCNN architectures for both test systems. While the improved GCNN models can reach a slightly lower MAE than the basic GCNN model for some setups, these improvements do not justify their increased model complexity and the subsequently higher computational requirements. For comparison, the basic GCNN model has around 710'000 tunable parameters, while the global GCNN model has around 1'200'000 parameters (870'000 in the interaction GCNN and 840'000 in the attention GCNN). In addition, the interaction GCNN and attention GCNN models require dot products of the query ($Q \in R^{n_{in}\times n_k}$) and key ($K \in R^{n_{in}\times n_k}$) matrices, which are computationally expensive (although the costs can be reduced through the use of multi-head self-attention and $n_{out} = n_k < n_{in}$). For these reasons, we decided to focus on the basic GCNN model and the $\Delta$-learning scheme in the following. 

\subsection{Model Parametrization} 
For the basic GCNN architecture, we explored different model parametrizations: edge-cutoff value ($r_\text{cutoff}$) from 0.3 to 0.6~nm, number of features per dense layer ($n_f$) from 32 to 256, and model depth (i.e., number of interaction blocks) from 2 to 8. The edge-cutoff value and the model depth directly influence the contributions from long-range interactions, as they both determine (directly and indirectly) the distance up to which atoms contribute to the message update function. The number of features per dense layer, on the other hand, determines the maximum possible complexity of each message update function.

\begin{figure}[H]
  \includegraphics[width=\textwidth]{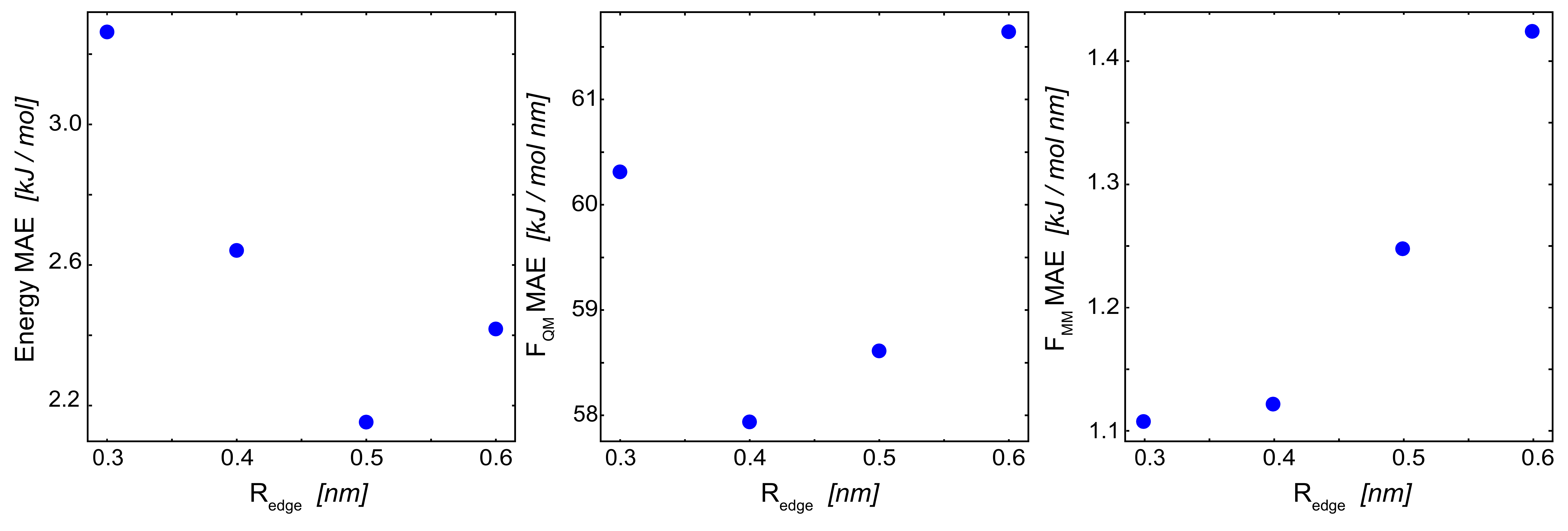}
  \caption{Influence of $R_\text{edge}$ on the mean absolute error (MAE) for the $\Delta$-learning basic GCNN model for uracil in water. (\textbf{Left}): MAE of the energies in the QM zone. (\textbf{Middle}): MAE of the forces on the QM particles. (\textbf{Right}): MAE of the forces on the MM particles from the QM zone. The numerical values are given in Table \ref{tbl:cutoff}.}
  \label{fgr:cutoff}
\end{figure}

Figure \ref{fgr:cutoff} shows the influence of the edge-cutoff value ($R_\text{edge}$) on the model performance for the $\Delta$-learning GCNN model of uracil in water. For the energy, the error is lowest for a $R_\text{edge}$ of 0.5~nm. For both force terms ($F_{QM}$ and $F_{MM}$), however, a smaller $R_\text{edge}$ of 0.4~nm and 0.3~nm, respectively, gives the lowest errors. Note that for both the energy and the force terms, the error increases again for larger $R_\text{edge}$ values. The reason for this could be that larger cutoffs are connected with an increase in possible system configurations, requiring in turn more training data points. If the same training set is used, overfitting might occur. For benzene in water, a different trend is observed. Here, a large $R_\text{edge}$ of 0.6~nm still leads to an improvement in the MAE for the forces (Table \ref{tbl:cutoff}). This is especially interesting as a different MM cutoff radius is used in the reference QM/MM calculations of benzene and uracil ($R_c =$0.6~nm and 1.4~nm, respectively). A reason could be, that the reduced MM cutoff radius ($R_c$) leads to a smaller configuration space in the MM-region and thus a GCNN model with a larger $R_\text{edge}$ is able to sufficiently learn the important contributions. 

\begin{table}[H]
  \caption{Influence of $R_\text{edge}$ on the mean absolute error (MAE) on the validation set for the basic GCNN models for the test systems benzene in water and uracil in water. For each property, the model with the lowest MAE is marked in bold.}
  \label{tbl:cutoff}
  \small
  \centering
  \begin{tabular}{l | rrr}
    \hline
    \multicolumn{4}{l}{\textbf{Benzene}} \\ \hline
    & \multicolumn{3}{c}{\textbf{$\Delta-$learning}} \\
    $R_\text{edge}$ & $E $  & $F_{QM} $& $F_{MM} $ \\
   nm & kJ~mol$^{-1}$  & kJ~mol$^{-1}$~nm$^{-1}$ & kJ~mol$^{-1}$~nm$^{-1}$ \\
    \hline
    0.3 & 2.4 & 27.8 & 5.5 \\
    0.4 & 1.3 & 23.7 & 4.2 \\
    0.5 & \textbf{1.0} & 22.9 & 3.5 \\
    0.6 & 1.1 & \textbf{21.0} & \textbf{3.1} \\
    \hline
    \hline
    \multicolumn{4}{l}{\textbf{Uracil}} \\ \hline
    &\multicolumn{3}{c}{\textbf{$\Delta-$learning}} \\
    $R_\text{edge}$ & $E $  & $F_{QM} $& $F_{MM} $ \\
   nm & kJ~mol$^{-1}$  & kJ~mol$^{-1}$~nm$^{-1}$ & kJ~mol$^{-1}$~nm$^{-1}$ \\
    \hline
    0.3 & 3.3 & 60.4 & \textbf{1.1}\\
    0.4 & 2.6 & \textbf{59.2} & \textbf{1.1} \\
    0.5 & \textbf{2.2} & 60.2 & 1.3 \\
    0.6 & 2.4 & 62.9 & 1.5 \\
    \hline
  \end{tabular}
\end{table}

For the model depth, we observe a similar trend as for the edge-cutoff value. Namely that including information about further distant nodes (atoms) is not always beneficial for the model performance (Table S1 in the Supporting Information). For the test system uracil in water, the MAE of both the energies and forces is lowest at around 2-4 interaction layers, while up to 6-8 interaction layers were required for the benzene in water test system. 
The results for the number of features per dense layer ($n_f$) are given in Table S2 in the Supporting Information. In general, a higher number of features per dense layer does not lead to a lower error, with convergence at around 64 features per dense layer for uracil in water. For the test system of benzene in water, on the other hand, a more complex model leads again to a decrease in the errors. 

\subsection{Loss Contribution of the Forces}
B\"oselt \textit{et al.}\cite{Boselt2021MachineSystems} have shown that the relative weighting of the different loss terms (energy loss, $F_{QM}$ loss, and $F_{MM}$ loss) can have a significant influence on the prediction accuracy. Thus, we systematically trained and evaluated GCNN models by varying the relative loss weightings ($wE, wF_{QM}$ and $wF_{MM}$). We decided to keep $wE$ constant at 1.0, while changing $F_{QM}$ from 0.001 to 100 and $F_{MM}$ from 0.1 to 1000. 

\begin{figure}[H]
  \centering
  \includegraphics[width=1.0\textwidth]{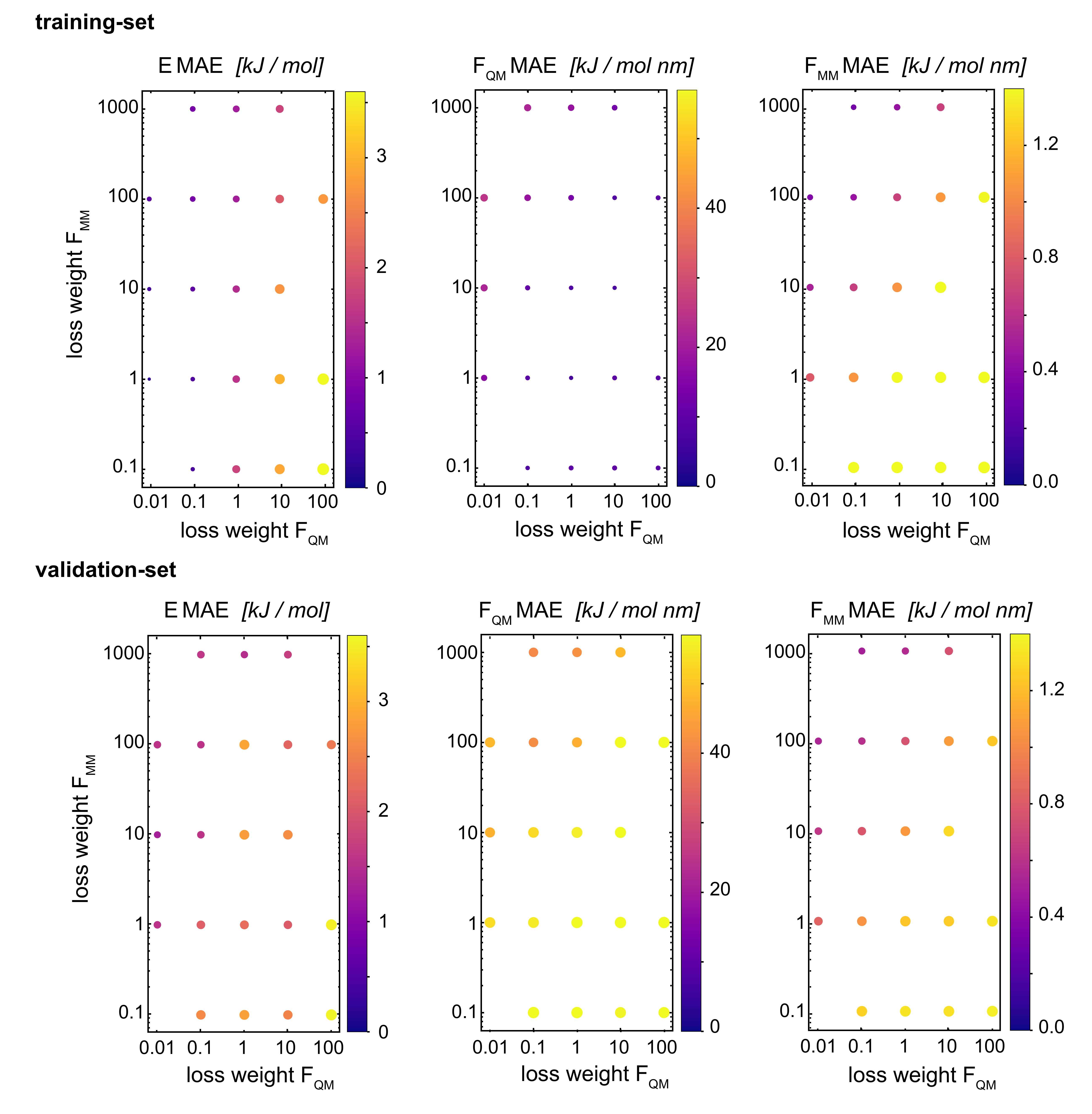}
  \caption{Influence of relative weights ($wF_{QM}$ and $wF_{MM}$) for the different loss terms (QM forces, MM forces) on the MAE of the energy (\textbf{left}), forces on QM particles (\textbf{middle}), and the forces on the MM particles from the QM zone (\textbf{right}). The basic GCNN with the $\Delta$-learning scheme and the test system uracil in water was used. The weight of the energy loss ($wE$) is kept constant at 1.0. (\textbf{Top}): MAE for the training set. (\textbf{Bottom}): MAE for the validation set.  The color map and the relative size of the points indicate the MAE.}
  \label{fgr:loss_uracil_delta}
\end{figure}

\begin{figure}[H]
  \centering
  \includegraphics[width=0.8\textwidth]{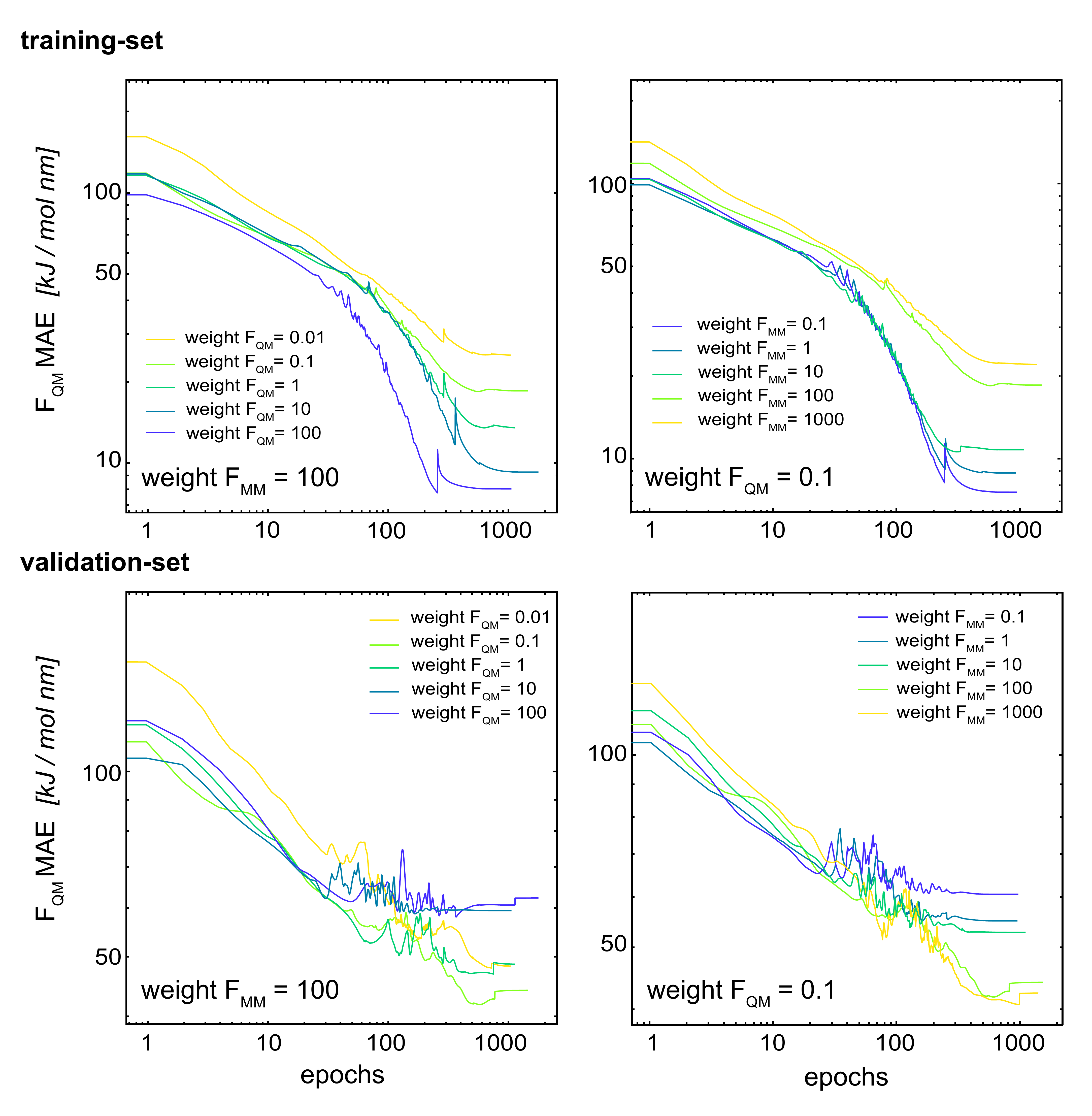}
  \caption{Learning curves for $F_{QM}$ in the training set (\textbf{top}) and validation set (\textbf{bottom}) when varying the relative loss weightings $wF_{QM}$ (\textbf{left}) and $wF_{MM}$ (\textbf{right}). The basic GCNN with the $\Delta$-learning scheme and the test system uracil in water was used. The weight of the energy loss ($wE$) is kept constant at 1.0.}
  \label{fgr:loss_lc_uracil_delta}
\end{figure}

Figure \ref{fgr:loss_uracil_delta} shows the MAE on the training and validation set for the basic GCNN model with the $\Delta$-learning scheme for the test system of uracil in water when varying $wF_{QM}$ and $wF_{MM}$. For the training set, we observe the expected behaviour, i.e. the error on $F_{QM}$ decreases when the weight $wF_{QM}$ is increased and the same for $F_{MM}$ with $wF_{MM}$. Interestingly, the results on the validation set are different. While the error on $F_{MM}$ still decreases with increasing $wF_{MM}$ as expected (bottom right panel in Figure \ref{fgr:loss_uracil_delta}), the MAE on $F_{QM}$ is smallest for low $wF_{QM}$ and high $wF_{MM}$ (bottom middle panel in Figure \ref{fgr:loss_uracil_delta}). This observation is further illustrated in Figure \ref{fgr:loss_lc_uracil_delta}, which shows the learning curves for $F_{QM}$ in the training and validation sets (corresponding to the horizontal line at $wF_{MM}=100$ and the vertical line at $wF_{QM}=0.1$ in the bottom middle panel in Figure \ref{fgr:loss_uracil_delta}). 

For the training set, the learning curves show again the expected dependence with $F_{QM}$ and $F_{MM}$, respectively (top panels in Figure \ref{fgr:loss_lc_uracil_delta}). However, for the validation set, the lowest error for $F_{QM}$ is observed with $wF_{QM}=0.1$ and not with $wF_{QM}=100$. In general, a relative weighting of $\frac{F_{MM}}{F_{QM}}$ around 100-1000 results in the lowest MAE on $F_{QM}$. 
Higher relative $wF_{QM}$ values seem to result in over-fitting of $F_{QM}$. Thus, for the final GCNN model, we use the following relative loss weights: $wE=1, wF_{QM}=0.1$ and $wF_{MM}=10$.
Note that the same observations were made for the $\Delta$-learning GCNN model with benzene in water (Figures S1 and S2 in the Supporting Information). 

\subsection{Neighborhood Reduction}
The main increase in computational requirements for the GCNN models (especially memory requirements when using GPUs during the batched training procedure) comes from the growing number of edges with an increasing size of the edge-cutoff ($r_\text{cutoff}$). For example, $r_\text{cutoff}=0.4$~nm leads to around 21'000 edges for one of the uracil in water snapshots, while $r_\text{cutoff}=0.5$~nm already leads to around 38'000 edges. 
Therefore, we investigated whether the number of edges can be reduced using different neighborhood schemes, without decreasing the model performance. To limit the number of edges around each atom, we explored the k-nearest-neighbours \cite{Fix1989DiscriminatoryProperties} method with eight (KNN-8) and twelve (KNN-12) neighbours, as well as the Voronoi–Dirichlet polyhedra (VD) \cite{NiggliXXIV.I., Blatov2004Voronoi-DirichletApplications}. With the same $r_\text{cutoff}=0.5$~nm, KNN-8 leads to around 12'000 edges for a uracil in water snapshot, KNN-12 to around 18'000 edges, and VD to around 11'000 edges. Thus, the memory requirement for the GCNN model is drastically reduced by all of these approaches. 

\begin{figure}[H]
  \includegraphics[width=\textwidth]{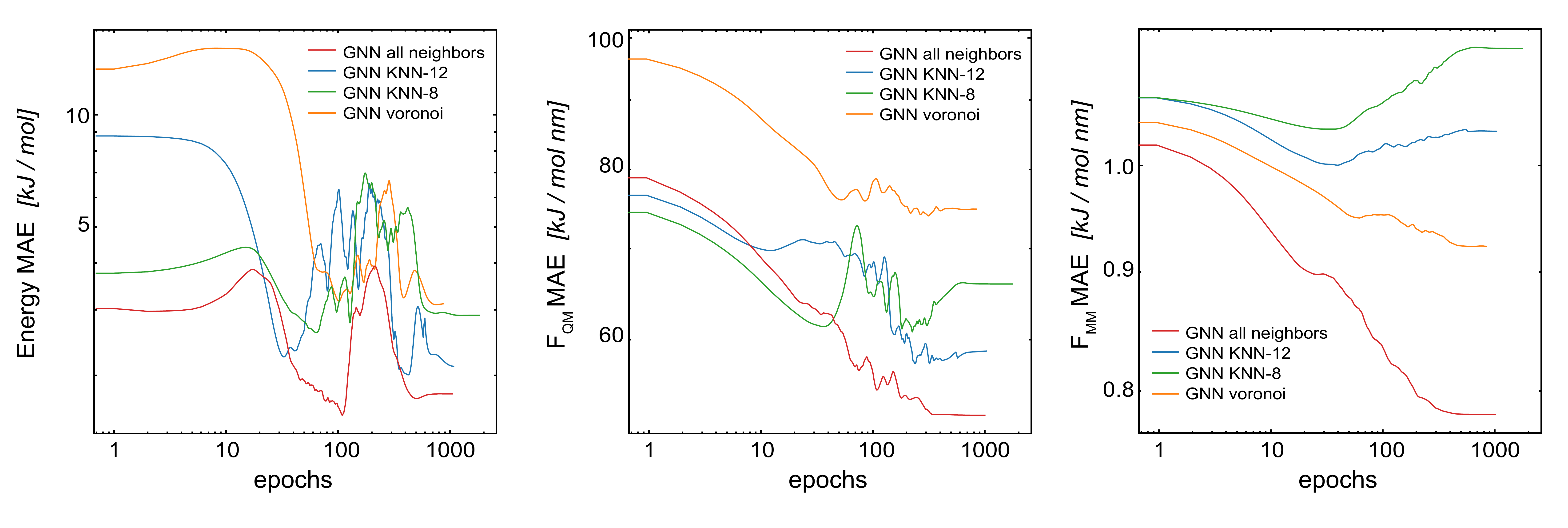}
  \caption{Learning curves on the validation set for the $\Delta$-learning GCNN model when using different neighborhood selection schemes. The test system uracil in water was used.}
  \label{fgr:NN_uracil}
\end{figure}

For the following results, we used the same settings for the GCNN models, i.e., $r_\text{cutoff}=0.5$~nm, $n_f=128$ with five interaction layers, and loss weightings of $wE=1.0, wF_{QM}=0.1$ and $wF_{MM}=10$.
Figure \ref{fgr:NN_uracil} shows the learning curves on the validation set for the $\Delta$-learning GCNN model for the test system uracil in water. While the MAE on the energies is largely unaffected (within the fluctuations of the error) by the choice of the the neighborhood scheme (KNN-12 being closest to the complete model), the errors on the force terms ($F_{QM}$ and $F_{MM}$) increase substantially with all neighborhood schemes compared to the complete model. 
The same trends are also observed for the $\Delta$-learning model of benzene in water (Table S3 in the Supporting Information). 

\subsection{Data Set Ordering}
The split into training/validation/test sets was chosen with the future application in MD simulations in mind. The idea is that a short initial QM/MM MD run of the target system can be used as the training set for subsequent longer (QM)ML/MM MD simulations. To mimic this, the first 70\% of the frame from the initial QM/MM trajectory were taken as training set, the following 20\% as validation set and the final 10\% as test set.\cite{Boselt2021MachineSystems} This leads to a time-based ordering of the frames within the training set. Here, we investigate if this correlated ordering has an effect on the model performance. For this, we compared five models trained with the same training set but using a different frame ordering within the set: (i) original time-based ordering from the MD simulation, (ii-iv) random-ordering using three different random number seeds, and (v) a farthest-point sampling (fps) as described in the Supporting Information. For all models, we used $r_\text{cutoff}=0.5$~nm, $n_f=128$ with five interaction layers, and loss weightings of $wE=1.0, wF_{QM}=0.1$ and $wF_{MM}=10$.

\begin{figure}[H]
  \includegraphics[width=\textwidth]{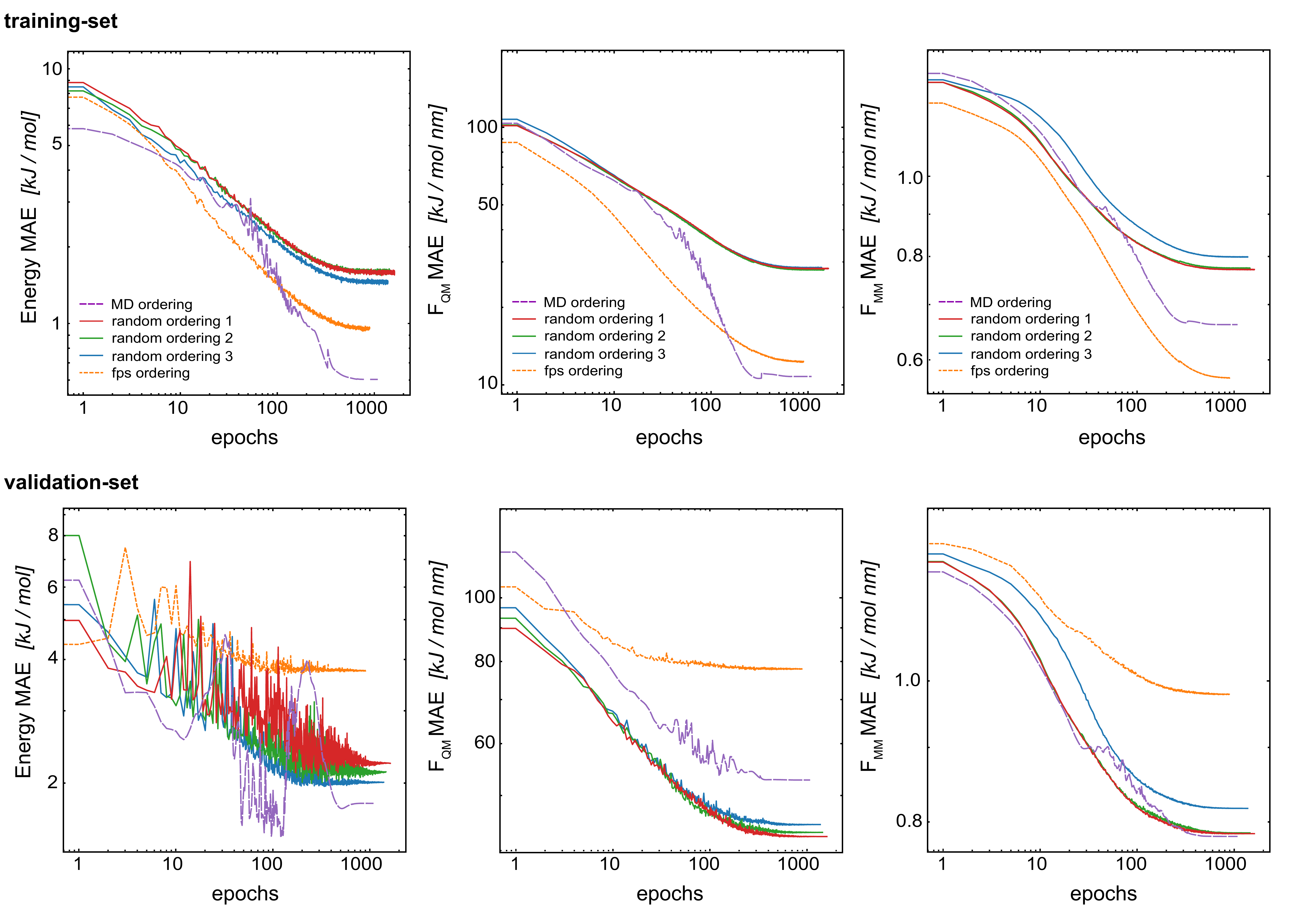}
  \caption{Learning curves for the training set (\textbf{top}) and validation set (\textbf{bottom}) when varying the order of the data points in the training set. The $\Delta$-learning GCNN for uracil in water is shown.}
  \label{fgr:fixed_delta}
\end{figure}

Figure \ref{fgr:fixed_delta} shows the learning curves for the training and validation set for the basic GCNN model with the $\Delta$-learning scheme trained with different orderings of the data points in the training set. First of all, the results clearly show that the order of the training data points does indeed affect the model performance. For all three properties (energies and forces), the error for the training set is clearly smallest with the fps-ordering (orange line) or MD-ordering (purple line). Interestingly, using an fps-ordering of the training data points results in a higher MAE for the validation set, indicating that the fps-ordering leads to an over-fitting of the model for this setup. The situation is different for the MD-ordering, which leads to comparable results on the validation set as the random-ordering, except for $F_{QM}$. Here, the accuracy in between the random and the fps-ordering.
For the test system of benzene in water (Figure S3 in the Supporting Information), 
the same trends are observed for fps-ordering on the training and validation set, while the MD-ordering shows again a comparable accuracy as the random-orderings for both the training and the validation set. However, in all cases the random-orderings converge earlier (at around 1000 epochs) compared to the MD-ordering (around 2000 epochs). Thus, using a random-ordering of the MD frames in the training set seems to be beneficial by reducing the training costs and potentially increasing robustness.

\subsection{Comparison to HDNNP}
As a final evaluation, we compare the GCNN models with the previously developed HDNNPs ,\cite{Behler2011Atom-centeredPotentials, Boselt2021MachineSystems} which use the same $\Delta$-learning scheme with DFTB \cite{Elstner2006TheSystems, Hourahine2020DFTB+Simulations} as baseline,  for all five test systems: (i) benzene in water, (ii) uracil in water, (iii) retionic acid in water, (iv) (close to) transition state of the $S_N2$ reaction of \ce{CH3Cl} with \ce{Cl-} in water, and (v) (close to) transition state of SAM with cytosine in water. The model performance is compared on the training set (7000 frames), validation set (2000 frames), and test set (1000 frames). Note that the models were solely developed with the training and validation sets, and are only at this stage evaluated on the test set. Furthermore, the hyperparameters were only tuned on the two test systems benzene and uracil in water. 
For all models, we used $R_\text{edge}=0.5$~nm, $n_f=128$ with four (systems i and ii) or five (systems iii-v) interaction layers, and loss weightings of $wE=1.0, wF_{QM}=0.1$ and $wF_{MM}=10$.

\begin{figure}[H]
  \includegraphics[width=\textwidth]{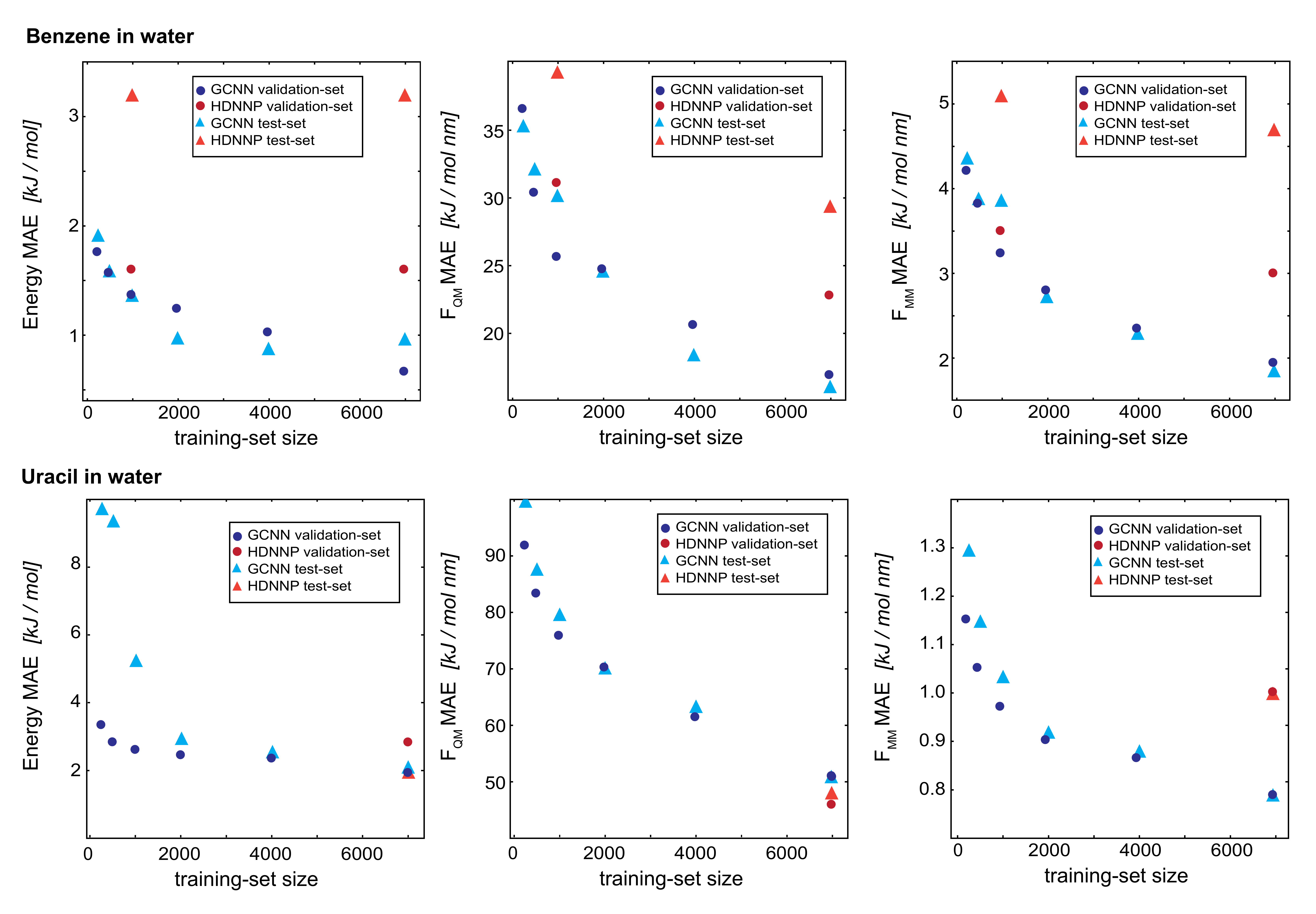}
  \caption{Learning curves as a function of the training-set size for the $\Delta$-learning models for benzene in water (\textbf{top}) and uracil in water (\textbf{bottom}). The dark blue circles show the GCNN validation MAE, the dark red circles show the HDNNP validation MAE, the light blue triangles show the GCNN test MAE, and the light red triangles show the HDNN test MAE. The numerical values are given in Table \ref{tbl:test_set}.}
  \label{fgr:uracil_lc}
\end{figure}

Figure \ref{fgr:uracil_lc} compares the learning curves and final MAE between the HDNNP and the GCNN models for benzene in (top panels) and uracil in water (bottom panels). For both systems, the learning curves for the GCNN validation and test set show a very similar performance. This indicates that the models are not overfitted and are able to generalize to new data points within the same MD trajectory. 
For the HDNNP models, we observe a similar behaviour. However, the MAE values for the test set of benzene in water are consistently above the MAE for the validation set. For this system, we can directly compare the effect of the training set size between the HDNNP and GCNN models. While both models show a similar MAE for small training set sizes, the GCNN model outperforms the HDNNP model when all training data points are used. A possible reason for this observation is that the GCNN model has to learn the descriptor for the atomic environments and would thus profit from a larger training set.
This is also indicated by the fact that the learning curves for the GCNN models are still continuously decreasing even at 7000 training structures. For the system of uracil in water, the final MAE values are comparable for all models. Here we note that the benzene test system has a shorter MM-cutoff radius ($R_c$), which might simplify the learning task and thus a similar behaviour might be observed for uracil at a larger training set size. 

Table \ref{tbl:test_set} shows the MAE values on the training, validation, and test set for all five test systems. 
Overall, the GCNN models perform similar or slightly better than the HDNNP models. A major exception to this is the SAM/cyt in water system. Here, the GCNN model reaches the same accuracy as the HDNNP model on the training set, but the MAE on the validation and test set is two to fives times higher than with the HDNNP model. This may be because the GCNN models were parametrized and benchmarked on the simpler systems (i.e., benzene and uracil in water) and there is less transferability of the hyperparameters than with the HDNNP.

\begin{table}[H]
  \caption{Mean absolute error (MAE) on the training set (7000 frames), validation set (2000 frames), and test set (1000 frames) for the $\Delta$-learning HDNNP models and the $\Delta$-learning GCNN models for all five test systems: (i) benzene in water, (ii) uracil in water, (iii) retionic acid in water, (iv) (close to) transition state of the $S_N2$ reaction of \ce{CH3Cl} with \ce{Cl-} in water, and (v) (close to) transition state of SAM with cytosine in water. For each test system, the model with the lowest MAE is marked in bold.}
  \label{tbl:test_set}
  \scriptsize
 \centering
  \begin{tabular}{l | rrr | rrr}
    \hline
    & \multicolumn{3}{c|}{\textbf{HDNNP}} & \multicolumn{3}{c}{\textbf{GCNN}} \\
     & $E $  & $F_{QM} $ & $F_{MM} $ & $E $  & $F_{QM} $& $F_{MM} $ \\
    System & kJ~mol$^{-1}$  & kJ~mol$^{-1}$~nm$^{-1}$ & kJ~mol$^{-1}$~nm$^{-1}$ & kJ~mol$^{-1}$  & kJ~mol$^{-1}$~nm$^{-1}$ & kJ~mol$^{-1}$~nm$^{-1}$ \\
    \hline
    Benzene & 1.8/1.6/3.2 & 23.7/22.8/29.4 & 3.5/3.0/4.7 & \textbf{0.4}/\textbf{0.7}/\textbf{0.7} & \textbf{8.6}/\textbf{14.7}/\textbf{15.3} & \textbf{1.3}/\textbf{1.9}/\textbf{2.0}\\
    Uracil & 1.2/2.8/\textbf{2.0} & 34.3/\textbf{45.9}/\textbf{48.1} & 1.1/1.0/1.0 & \textbf{0.8}/\textbf{1.9}/2.1 & \textbf{14.1}/51.0/51.0 & \textbf{0.7}/\textbf{0.8}/\textbf{0.8}\\
    Chloroform &  1.8/1.7/2.2 & 24.9/29.5/30.7 & 1.0/1.0/1.0 & \textbf{0.5}/\textbf{1.1}/\textbf{1.2} & \textbf{5.3}/\textbf{21.4}/\textbf{22.1} & \textbf{0.5}/\textbf{0.6}/\textbf{0.6}\\
    Retionic acid & 3.9/\textbf{4.4}/- & 43.8/44.8/- & 1.1/1.1/- & \textbf{0.6}/\textbf{4.4}/\textbf{4.4} & \textbf{20.5}/\textbf{37.0}/\textbf{37.0} & \textbf{0.7}/\textbf{0.9}/\textbf{0.9}\\
    SAM/cyt & 6.3/\textbf{8.5}/- & \textbf{74.8}/\textbf{74.6}/- &  \textbf{2.3}/\textbf{2.3}/-& \textbf{5.6}/21.1/28.8 & 79.7/130.4/130.4 & 11.9/12.0/12.2\\
    \hline
  \end{tabular}
\end{table}

Up to this point, we have evaluated the performance of the GCNN models in terms of MAE on a reasonable validation  and test set. B\"oselt \textit{et al.}\cite{Boselt2021MachineSystems} have already emphasized that it is important to test a ML model for the intended application, which in our case are (QM)ML/MM MD simulations. While rare outliers get averaged in the MAE assessment, they might be less tolerable in an actual simulation, where the results of the next step depend directly on the results of the previous step. For this reason, it is crucial to test the performance of the developed (QM)ML/MM models in a prospective MD simulation. 
We performed therefore (QM)ML/MM MD simulations using the $\Delta$-learning GCNN model for the two test systems with larger conformational flexibility as in Ref.~\citenum{Boselt2021MachineSystems}: (i) retinoic acid, and (ii) (close to) transition state of SAM with cytosine in water. 
Note that the models were only trained on the initial 7'000 steps of the QM/MM MD simulations and no adaptive re-training during the (QM)ML/MM production runs was performed. Shen and Yang\cite{Shen2018MolecularNetworks} have shown that adaptive neural networks can use on-the-fly corrections of the model to further improve the model performance. However, this comes with an increase in the training cost of the model and changing energies/forces (i.e., the estimated properties of a configuration may not be the same at the beginning versus the end of the simulation).

\begin{figure}[H]
\centering
  \includegraphics[width=0.8\textwidth]{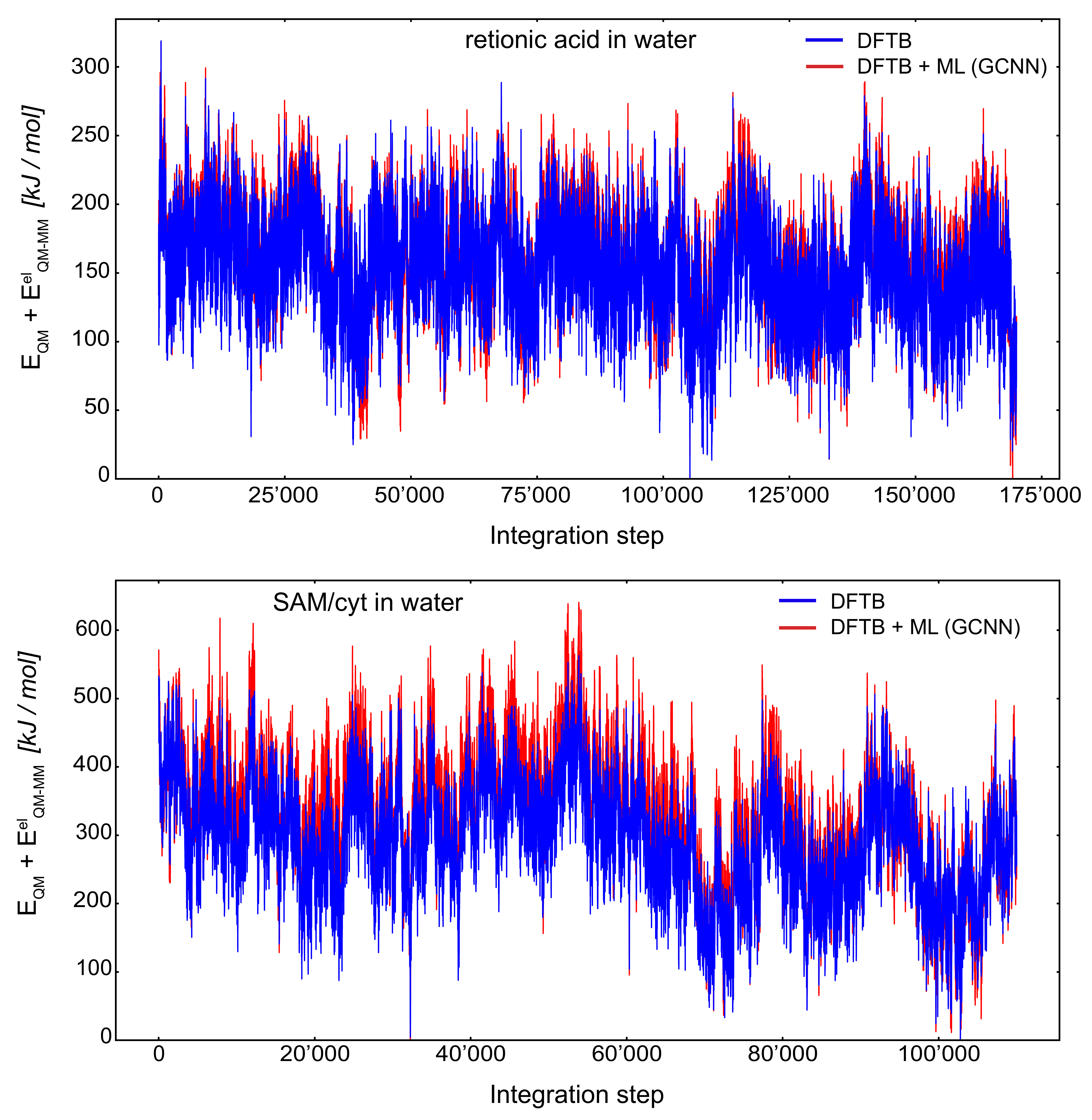}
  \caption{MD simulation of retionic acid in water (top) and SAM/cyt in water (bottom) using an integration time step of 0.5 fs. The energy ($E_{QM} + E^{el}_{QM-MM}$) trajectory is extracted from 170'000 (retionic acid) and 110'000 (SAM/cyt) consecutive steps performed by the DFTB + GCNN model. The first 200 steps were discarded as equilibration. The pure DFTB energy is shown in blue and the GCNN $\Delta-$correction in red. Note that the GCNN model was only trained on the initial 7'000 QM/MM MD simulation steps (not shown here) and that no adaptive on-the-fly re-training of the GCNN model was performed.}
  \label{fgr:MD}
\end{figure}

Figure \ref{fgr:MD} shows the MD energy ($E_{QM} + E^{el}_{QM-MM}$) trajectories for retionic acid in water (top panel) and SAM/cyt in water (bottom panel) using an integration step of 0.5 fs. The (QM)ML/MM MD simulation of retionic acid in water was carried out for 170'000 consecutive steps, while the simulation of SAM/cyt in water was performed for 110'000 consecutive steps. For both systems, the first 200 steps were discarded as equilibration.
Thus, it was possible to propagate these systems stably using the $\Delta-$learning GCNN model based on a fraction of the otherwise required QM/MM MD simulation steps. For both systems, the energy fluctuations during the (QM)ML/MM MD simulation were comparable to the ones with the $\Delta-$learning HDNNP model: For retionic acid, $\sigma_{DFTB+GCNN}=37.1 kJ/mol$ and $\sigma_{DFTB+HDNNP}=55.2 kJ/mol$, and for SAM/cyt, $\sigma_{DFTB+GCNN}=94.9 kJ/mol$ and $\sigma_{DFTB+HDNNP}=73.3 kJ/mol$. Of course, the possibility that in a longer simulation an ill-represented structure may be encountered cannot be fully excluded.
Interestingly, even though the accuracy of the GCNN model on the validation/test sets of the SAM/cyt in water was considerably lower than with the HDNNP model, we still observe a stable and robust simulation over 110'000 (QM)ML/MM MD steps based solely on the initial 7'000 QM/MM MD simulation steps.  

\section{Conclusion}
In this work, we investigated the use of GCNN models in (QM)ML/MM MD simulations for condensed-phase systems with DFT accuracy for the QM subsystem, and compared the results to the previously developed HDNNPs for the same systems.
In a first step, we evaluated different GCNN architectures capable of incorporating long-range effects, with and without a $\Delta$-learning scheme. While no large improvements could be found with the more complex GCNN architectures, the $\Delta$-learning GCNN using DFTB as baseline yielded the most accurate description of the energies and forces. Based on these observations, we focused on the basic GCNN with $\Delta$-learning for the remainder of the study. Next, we assessed the influence of different parameters on the performance of the GCNN models. Here we found that the inclusion and the correct weighting of the QM and MM gradients in the loss function were crucial to improve model performance. Interestingly, we observed that increasing the loss weights for the QM gradients does not lead to an improved accuracy when predicting them due to overfitting. Instead, relative loss weights of $wE=1, wF_{QM}=0.1$ and $wF_{MM}=10$ provided the model with the best ability to generalize to new data points.

In order to reduce the computational requirements of the GCNN models, we also investigated different neighborhood reduction schemes in the creation of the GCNN edges. While these schemes decrease indeed the computational costs without significantly affecting the accuracy of the QM energies,  the performance on the QM and MM forces is clearly worse. Thus, we do not recommend to use these schemes in (QM)ML/MM MD simulations, as predicted forces are directly used to propagate the system in time.

An interesting observation was made regarding the order of the data points in the training set.
In order to mimic the future application, the training set was chosen as the first 70\% of frames from a QM/MM trajectory. This lead to a time-based ordering of the training data points. When using a random ordering of the training set (without changing the actual training/validation/test split!) a similar model performance was reached, but the random ordering converges faster and thus leads to a reduction in the required training time.

Finally, we compared the $\Delta$-learning HDNNP and $\Delta$-learning GCNN models with each other for five different test systems in water. While both models perform at a similar accuracy, the GCNN model reaches slightly lower MAE values for most of the five test systems. The $\Delta$-learning GCNN model can also be used to perform stable (QM)/ML/MM MD simulations as the corresponding HDNNP model. However, the two model types differ drastically in their architecture and come with advantages and disadvantages, which should be considered when choosing an adequate model for a (QM)ML/MM simulation.
The symmetry functions in HDNNPs include a cosine term for all the inter-atomic angles within an atomic environment, which results in a computational scaling of $\mathcal{O}(N^3)$, where $N$ is the combined number of atoms in the QM and MM environments. The GCNN model, on the other hand, only depends on the edge-update operations to describe the atomic environments resulting in a computational scaling of $\mathcal{O}(N^2)$. 
Note that by using a finite cutoff for the atomic environments and edge-update operations, the scaling can be reduced to $\mathcal{O}(N (\log N)^2)$ and $\mathcal{O}(N \log N)$, respectively.
Additionally, the number of angle terms in the HDNNP symmetry functions scale exponentially with the number of element types (which can be partially resolved by introducing weighted symmetry functions), whereas the GCNN atom types are encoded directly in an embedding vector and do not directly change the scaling of the model.
Another important difference is that HDNNPs are based upon individual neural-network potentials (NNP) for each atomic environment, whose evaluation can be easily distributed over multiple GPUs and the memory requirement for each individual NNP does not increase drastically with  increasing QM or MM system size. However, In contrast, GCNNs are based upon iterative and overlapping message passing operations and thus, a distributed evaluation is not as straightforward. Additionally, for each message passing operation within the GCNN graph, all edges have to be transformed using a dense layer. Therefore, the memory requirement for a GCNN evaluation scales as $\mathcal{O}(N_e \cdot n_f)$, where $N_e$ is the number of edges. This means that the memory increases with increasing QM or MM system size. Note that this can be critical when using a batched training procedure on a GPU.

In summary, the HDNNP based $\Delta$-learning (QM)ML/MM setup appear to be best suited for condensed-phase systems with a limited number of different element types in the QM zone and the surrounding MM zone. Examples are reactions of organic molecules in a solvent -- similar to the five test systems investigated in this paper -- or reactions at the interface of a mono-atomic surface. In these cases, the limited number of symmetry functions and the parallelizability over multiple GPUs / CPUs favor the HDNNP setup. On the other hand, the GCNN based $\Delta$-learning (QM)ML/MM setup is most useful for condensed-phase systems with a larger number of element types, as the model scaling of GCNN is much less affected by the number of element types than HDNNP. Examples are reactions within the active site of a metalloenzyme, proteins with covalently bound or interacting ligands, reactions involving organometallic catalysts, or reactions catalyzed at the interface of a poly-atomic surface.

\section*{Acknowledgements}
The authors thank Moritz Th\"urlemann for the helpful discussions about GCNNs and HDNNPs, and Felix Pultar for his help in debugging the ML/MM code and the insightful discussions.
The calculations were carried out on the high-performance cluster Euler of ETH Zurich.

\bibliography{references} 

\begin{thebibliography}{10}
\bibitem{Lin2007QM/MM:Here}
Hai Lin and Donald~G. Truhlar.
\newblock {QM/MM}: {W}hat have we learned, where are we, and where do we go
  from here?
\newblock {\em Theo. Chem. Acc.}, 117:185--199, 2007.
\end{thebibliography}
\bibliographystyle{unsrt} 

\end{document}